\documentclass[conference]{IEEEtran}
\IEEEoverridecommandlockouts
\usepackage{flushend}
\usepackage{caption}
\usepackage{subfigure}
\usepackage{subcaption}
\usepackage{graphicx}
\usepackage[table, svgnames, dvipsnames]{xcolor}
\usepackage{booktabs}
\usepackage{comment}
\usepackage{cite}
\usepackage{amsmath,amssymb,amsfonts}
\usepackage{algorithmic}
\definecolor{green}{rgb}{0.16, 0.67, 0.53}
\usepackage{textcomp,balance,url}
\usepackage[left=1.62cm,right=1.62cm,top=1.9cm]{geometry}
\definecolor{chestnut}{rgb}{0.97, 0.51, 0.47}
\def\BibTeX{{\rm B\kern-.05em{\sc i\kern-.025em b}\kern-.08em
    T\kern-.1667em\lower.7ex\hbox{E}\kern-.125emX}}
\begin{document}

\title{Performance Measurements in the AI-Centric Computing Continuum Systems}

\author{
\IEEEauthorblockN{Praveen Kumar Donta\IEEEauthorrefmark{1}, Qiyang Zhang\IEEEauthorrefmark{2} and Schahram Dustdar\IEEEauthorrefmark{3}}

\IEEEauthorblockA{\IEEEauthorrefmark{1}\textit{Department of Computer Systems and Sciences}, \textit{Stockholm University,} Stockholm 164 25, Sweden.} 
\IEEEauthorblockA{\IEEEauthorrefmark{2}\textit{Computer Science School}, \textit{Peking University,}Beijing 100876, China.} 
\IEEEauthorblockA{\IEEEauthorrefmark{3}\textit{Distributed Systems Group, TU Wien}, Vienna 1040, Austria and \textit{ICREA} Barcelona 08002, Spain}
 \texttt{praveen@dsv.su.se}, \texttt{qiyangzhang@pku.edu.cn}, \texttt{dustdar@dsg.tuwien.ac.at}
}

\maketitle
\thispagestyle{plain}
\pagestyle{plain}
\begin{abstract}
Over the Eight decades, computing paradigms have shifted from large, centralized systems to compact, distributed architectures, leading to the rise of the Distributed Computing Continuum (DCC). In this model, multiple layers such as cloud, edge, Internet of Things (IoT), and mobile platforms work together to support a wide range of applications. Recently, the emergence of Generative AI and large language models has further intensified the demand for computational resources across this continuum. Although traditional performance metrics have provided a solid foundation, they need to be revisited and expanded to keep pace with changing computational demands and application requirements. Accurate performance measurements benefit both system designers and users by supporting improvements in efficiency and promoting alignment with system goals. In this context, we review commonly used metrics in DCC and IoT environments. We also discuss emerging performance dimensions that address evolving computing needs, such as sustainability, energy efficiency, and system observability. We also outline criteria and considerations for selecting appropriate metrics, aiming to inspire future research and development in this critical area.
\end{abstract}


\section{Introduction}\label{sec:Introduction}
\IEEEPARstart{C}{omputing} paradigms have evolved significantly over the past few decades, moving from large, room-sized processors and memory units to compact and efficient computing nodes. At the same time, system architectures have shifted from centralized models to distributed frameworks. One of the important developments in this direction is the Distributed Computing Continuum (DCC), which reflects a major change in how computing systems are designed and utilized \cite{lapkovskis2025benchmarking}. In contrast to earlier models, where data centers, cloud platforms, edge devices, and Internet of Things (IoT) systems operated independently, the DCC brings these layers together into a unified and collaborative environment \cite{donta2024human_based,donta2023exploring,pujol2023edge}. This integration enables applications to choose the most suitable layer for executing tasks or processing data. Depending on the requirement, processing can take place in a central data center, on cloud infrastructure, at the network edge, or directly on resource-constrained devices such as sensors or IoT nodes \cite{casamayor2022distributed}. In many cases, applications may also distribute their workloads across multiple layers at the same time to utilize resources efficiently and achieve highest quality of service (QoS).

\begin{figure}[!t]
    \centering
    \includegraphics[width=0.47\textwidth]{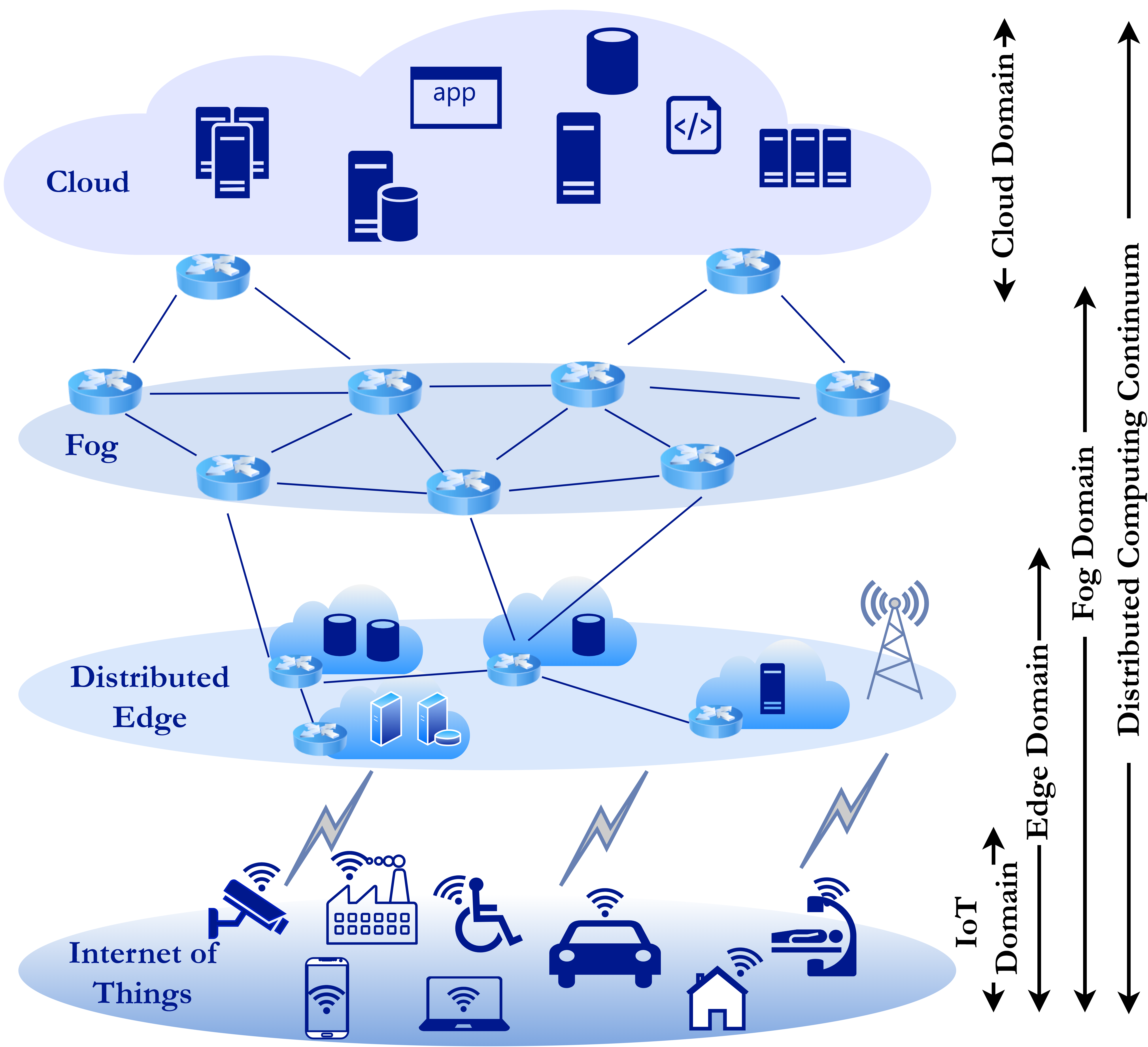}
    \caption{Conceptual Framework of DCC Architecture}
    \label{fig:dcc}
\end{figure}

On the other side of this evolution, the emergence of Generative AI (GenAI) and large language models (LLMs) has introduced new dimensions to computing \cite{mangione2025generative,alaa2025Towards,morabito2025smaller}. These models require significantly higher computational power compared to traditional applications, and their growing adoption has placed unprecedented demands on underlying infrastructures. As a result, there has been a noticeable shift towards high-performance computing systems, supported by advanced processors, specialized Graphics processing units (GPU), accelerators, and optimized memory and storage units \cite{silvano2025survey}. Over the past few years, global investment in AI hardware has steadily increased~\cite{ha2024investigation}, with leading data centers and research institutions upgrading their systems to handle larger workloads and improve throughput.

However, these technological advancements also highlight a growing mismatch between modern application demands and traditional performance measurement strategies \cite{heidelberger1984computer,calingaert1967system}. Most conventional metrics were developed with static, centralized systems in mind and often fall short in capturing the complexity, scale, and heterogeneity of today’s distributed environments \cite{wang2024performance}. This challenge becomes even more relevant in environments like the DCC, where powerful servers must work in coordination with lightweight, resource-constrained IoT devices. The wide variation in processing power, network reliability, and energy limitations across these layers makes it difficult to use a single method to assess system performance. Therefore, it is important to revisit existing evaluation strategies and design performance metrics that reflect the continuous evolution of computing systems and the growing demands of modern applications. 

In this context, we examine commonly used performance metrics and highlight where they may fall short in meeting the needs of today’s diverse and changing computing environments. We also suggest new directions for performance measurement that are better suited to modern infrastructure and emerging application demands. In summary, the main contributions of this paper are as follows:
\begin{itemize}   
    \item We explore traditional performance metrics at the computing level, at the network level, and at the application or user level. 
    \item We provide possible novel performance measurement approaches for the changing computational demands 
     \item We discuss the criteria to choose performance metrics according to their goals and characteristics regarding the available resources and applications demands. 
\end{itemize}

% The remaining sections of this paper are organized as follows. Section \ref{sec2} highlights the importance of performance measurement in new and evolving systems. Section \ref{sec3} reviews commonly used performance metrics. Section \ref{sec4} presents novel performance metrics for changing computing demands and discusses criteria for selecting appropriate metrics. Finally, Section \ref{sec5} concludes the paper.

\section{Importance of Performance Measurements}\label{sec2}
Measuring performance of computing systems is crucial for various reasons and this holds equally true for DCC. Firstly, it aids in the selection and architecture of computer systems by allowing practitioners to make informed choices based on performance and cost considerations. Performance evaluation also plays an important role in the design of systems and applications, enabling designers to choose configurations that offer the best performance. Moreover, it allows for the analysis of existing systems, identifying areas for improvement and optimization. In order to improve computing systems efficiency, it is essential to assess computer performance, which provides insights into component capacity, hardware, software, and operational procedures \cite{kant1992introduction}. In addition, performance evaluations help determine compatibility with various application requirements. 

As discussed earlier, DCC differ from regular computing systems in that they have distributed, interconnected, and heterogeneous computing nodes of different types connected over network. Due to these variations, measuring DCC performance is different compared to regular computing systems. Benchmarking and specific methods tailored to application domains must be used to diagnose a computing system's performance efficiently, this holds for DCC.

Systematic performance evaluation also enables stakeholders to make informed decisions regarding system design, deployment, and management. This process supports the identification of strengths and weaknesses within the system, highlights opportunities for improvement, and ensures that resources are utilized effectively. Moreover, regular performance assessment helps maintain the reliability, scalability, and adaptability of DCCS in the face of changing workloads, technologies, and user requirements. Ultimately, robust performance measurement fosters innovation and continuous improvement, enabling distributed systems to meet the diverse and evolving needs of modern applications.

\section{Traditional Performance Measurements}\label{sec3}
% This section begins by discussing the primary goals of performance evaluation in DCC, followed by a concise overview of commonly used metrics. 
\begin{figure*}[!t]
    \centering
    \includegraphics[width=0.82\textwidth]{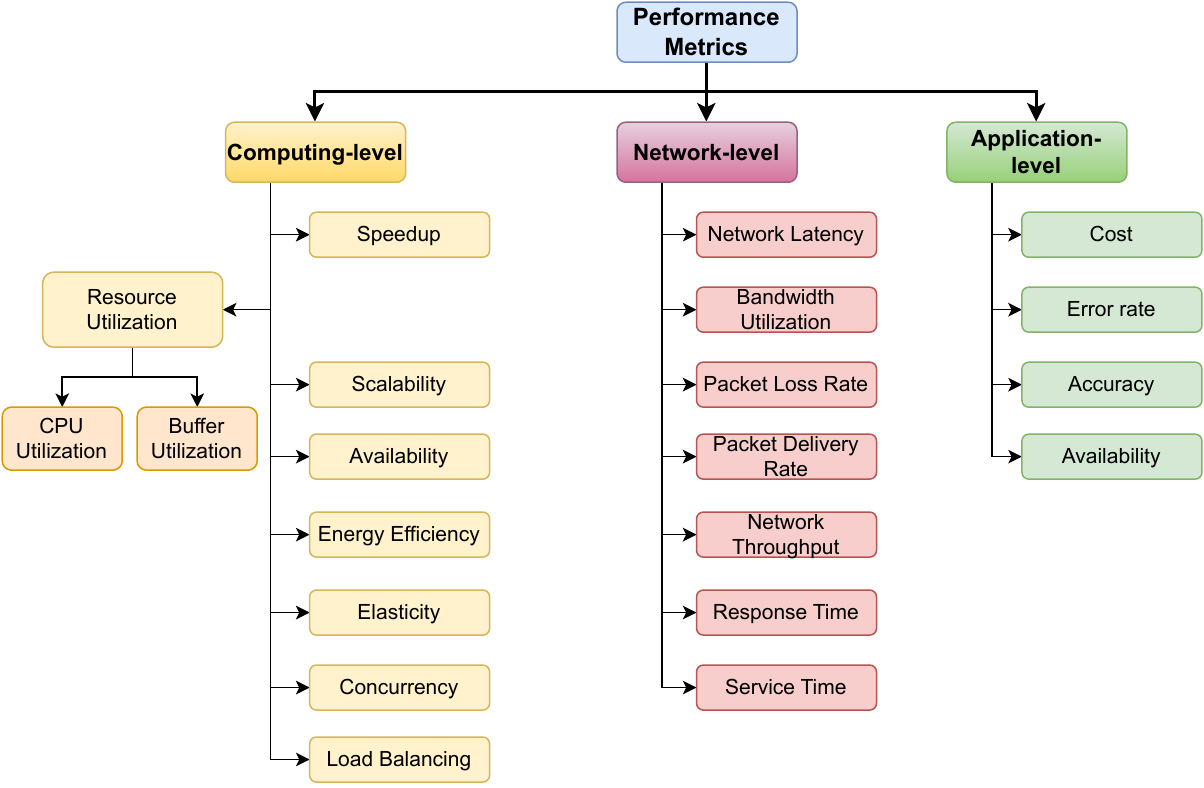}
    \caption{A taxonomy of commonly used  performance measurements in DCC and IoT Networks }
    \label{fig:taxonomy}
\end{figure*}
\subsection{Common goals of performance metrics in DCC}
%In this paragraph, I am trying to discuss the 'Common goals of performance metrics for DCCS'
Measuring DCC performance depends on the system’s operating conditions and its intended purpose. Still, several objectives remain consistently important for both developers and users. These include optimizing performance, improving scalability, maintaining QoS, identifying bottlenecks early, increasing energy efficiency, reducing costs, and strengthening reliability and security \cite{jain1990art}. Traditional performance optimization focuses on minimizing latency and maximizing resource utilization across the system. Scalability involves adapting to varying workloads without compromising performance, while QoS ensures that applications continue to meet expected service levels. In addition, system performance can be influenced by factors such as network latency, limited bandwidth, inefficient processing methods, or hardware constraints. Another crucial benefit of a complete measuring of DCC is to minimize its energy consumption. Most IoT devices use low-power batteries, while cloud devices with high computing power demand huge energy, all in all, results in large energy costs and carbon emissions, which affect the sustainability of DCC. 

The operational costs of DCC are not only limited to energy but also to computing and memory resources. A more efficient algorithm might be used to perform the high computations within limited resources to minimize the overall infrastructure costs. 
Reliability aims to ensure a high rate of recovery from failures and provide uninterrupted services. 
Security is an essential goal and it can continuously identify vulnerabilities or security threats in the system \cite{lilja2005measuring}. In general, detecting and addressing these issues early is essential to maintain stability, prevent degradation, and ensure long-term system effectiveness.

\subsection{Commonly used Performance Measurements}
Currently, researchers use a variety of performance metrics to evaluate DCC systems either as a whole or across specific layers, depending on the system designers. This paper categorizes these metrics into three groups: computing-level metrics, which capture system resource behavior; network-level metrics, which reflect communication efficiency; and application-level metrics, which assess user experience and data processing. The taxonomy of these performance metrics is illustrated in Fig.~\ref{fig:taxonomy} and discussed in detail below.

\subsubsection{Computing-level Performance Measurements}\label{sec:ComputingLevel} are essential for understanding how effectively computational resources are being utilized within a DCC and provide a comprehensive view of system behavior. These resources span across various components such as CPUs, memory, storage, and specialized hardware like GPUs or FPGAs distributed through cloud servers, edge devices, and IoT nodes. Measuring performance in such heterogeneous and dynamic settings requires a nuanced understanding of several critical factors. Metrics such as \textbf{speedup} provide insight into how much faster a task executes when distributed across multiple nodes compared to a single processor, helping to gauge the benefits of parallelism. \textbf{Resource utilization} reveals the extent to which the available hardware (CPU, Memory, other peripherals) is being used, highlighting potential over- or under-utilization of critical components. \textbf{Scalability} measures the system’s ability to handle increasing workloads by expanding resources, which is vital in scenarios where demand fluctuates rapidly. \textbf{Elasticity} captures how well the system adjusts resources in response to changing demands, maintaining cost-effectiveness and performance stability. \textbf{Availability} ensures that the system remains operational and accessible, an important aspect for applications requiring high reliability. In addition, \textbf{energy utilization} has become increasingly significant as the environmental impact and operational costs of large-scale deployments grow.  Finally, \textbf{concurrency} reflects the system’s capacity to perform multiple operations simultaneously to improve parallelism. %Collectively, these metrics provide a comprehensive view of system behavior, informing decisions about resource allocation, workload distribution, and energy management that ultimately improve performance and sustainability within the continuum.

\subsubsection{Network-level Performance Measurement}\label{sec:NetworkLevel}
evaluate the communication infrastructure to detect communication bottlenecks, fine-tune data routing, balance workloads (checking congestion), and maintain QoS. As cloud servers, edge devices, and IoT nodes in DCC depends on continuous data exchange to function cohesively, understanding network behavior becomes crucial. Since DCC is widely distributed and inherently heterogeneous environments, maintaining stable and efficient network performance  to ensuring reliable system operation. One of the core considerations is \textbf{network latency}, which refers to the time it takes for data to travel between two nodes.  \textbf{Bandwidth} represents the maximum data capacity of the network, while \textbf{throughput} reflects the actual volume of data successfully transferred over time. Sustained high throughput is vital for handling large-scale data and maintaining synchronization across distributed nodes. \textbf{Packet loss rate} captures the percentage of data packets that fail to reach their destination, potentially degrading both performance and reliability. In contrast, the \textbf{packet delivery rate} indicates the proportion of successfully received packets and serves as a measure of communication stability. \textbf{Network utilization} shows how much of the available capacity is in use, to monitor operational efficiency.

\subsubsection{Application or User-level Performance Measurement}\label{sec:ApplicationLevel}
focus on how effectively services and applications function from the perspective of the end user within a DCC. These metrics help evaluate service quality, reliability, and the system’s ability to meet specific application demands. \textbf{Response time}, which includes processing, queuing, and network delays, reflects how quickly the system responds to user requests and is key to ensuring smooth and interactive experiences. \textbf{Service time}, on the other hand, measures how long it takes to complete a given task, offering insight into the system's processing efficiency. \textbf{Cost} metrics are used to understand the economic impact of running applications, including the use of computing resources and overall operational overhead. \textbf{Reliability} is captured through the error rate, indicating how often applications fail or produce incorrect results, while accuracy measures the correctness of output. \textbf{Availability} tracks whether services are accessible when users need them.

\section{Novel Performance Measurements}\label{sec4}
% The following subsections explain novel metrics and the criteria for their selection.

\subsection{Metrics}\label{section41}
Many performance measures exist in the literature to evaluate computing systems. However, the continuous growth of these systems and the evolving demands of applications make it clear that new and more relevant evaluation methods are needed to accurately assess system performance.

\subsubsection{$CO_2$ emissions} GenAI and LLMs require large amounts of computing power, which leads to high electricity use and serious environmental impact. Training a single advanced model can consume as much electricity as several hundred homes use in a year, producing hundreds of tons of carbon ($CO_2$) emissions. After training, running these models for daily user requests continues to use significant computational power (which further use energy), with each query often consuming more power than a normal web search. Data centers that support these systems also need large volumes of water for cooling, which can affect local water availability. As demand increases, emissions from data centers may triple by 2030, potentially reaching 2.5 billion tons annually.  In addition, production of specialized hardware like GPUs further adds to the environmental impact. 

Just imagine the impact when IoT and edge devices begin integrating GenAI into their operations. With millions of such devices deployed worldwide—ranging from smart home assistants and wearable health monitors to industrial sensors and autonomous vehicles—the scale of energy consumption could increase dramatically. Unlike centralized data centers, these edge devices often operate in environments with limited power and cooling infrastructure, making energy efficiency a critical concern. Running even lightweight AI models locally requires significant processing power, which could lead to frequent battery drain, increased hardware wear, and added maintenance needs. Moreover, the cumulative energy demand of millions of devices performing AI tasks, even intermittently, could place additional pressure on electricity grids and raise global $CO_2$ emissions. Therefore, it is important to consider $CO_{2}$-related metrics when designing new system models.  

Vahdat et al.~\cite{vahdat2024new} proposed a performance evaluation metric to estimate carbon dioxide and equivalent emissions, referred to as goodput. Their approach incorporates established standards such as the Greenhouse Gas Protocol and ISO 14040 and 14044 to ensure accurate and reliable measurements. In the context of DCC and IoT, this metric is essential due to the rising importance of environmental responsibility and the drive toward sustainability.

\subsubsection{Heat Dissipation}
Similar to $CO_2$ emissions, the heat dissipation metric is closely linked to the environmental impact of DCC and IoT systems. Heat dissipation refers to the amount of heat produced and released by devices during their operation. This metric influences not only energy efficiency but also cooling needs and the overall sustainability of the system. To measure this metric, a popular approach, \textit{Newton's law of cooling} (shown in Eq.~(\ref{eq_Newton})), can be used \cite{winterton1999newton}. In this, the rate of heat loss of a device is proportional to the difference in temperature between the device and its environment. It helps in designing cooling strategies for data centers, edge devices, and IoT nodes by predicting temperature variations and optimizing cooling resources. 
% \begin{equation}\label{eq_Newton}
%     \frac{dT(t)}{dt} = -k \bigl(T(t) - T_{\text{env}}\bigr)
% \end{equation}
\begin{equation}\label{eq_Newton}
    T(t) = T_{\text{e}} + \bigl(T_0 - T_{\text{e}}\bigr) e^{-kt}
\end{equation}
where $T(t)$ is the device temperature at time $t$; $T_{\text{e}}$ denotes the ambient (environmental) temperature of the location where the device is deployed; $T_0$ represents the initial temperature of the device; $k$ is a positive constant that depends on the properties of the device and the cooling medium (i.e., it determines the rate of heat loss); and $e$ denotes Euler's number (approximately 2.718), which is the base of the natural logarithm.

\subsubsection{Bottlenecks}
In DCC, resource usage is often uneven. Some devices may be heavily loaded with tasks while others remain underutilized. Similarly, energy consumption can be much higher on certain nodes, and bandwidth availability may fluctuate across different parts of the network. These conditions create potential bottlenecks that are often hidden when relying on average metrics, as such measures tend to smooth out the underlying imbalances. To effectively identify and measure these hidden bottlenecks, the \textbf{fairness index} provides a valuable tool for assessing how evenly resources are distributed across the system. \textit{Jain’s fairness index} is one of the most popular metric to measure it \cite{jain1999throughput}, and  shown in Eq.~(\ref{eq_Jain}).
\begin{equation}\label{eq_Jain}
    \mathcal{F}(x_1, x_2, \ldots, x_n) = \frac{\left(\sum_{i=1}^n x_i\right)^2}{n \sum_{i=1}^n x_i^2}
\end{equation}
where $x_i$ represents the resource metric for the $i^{th}$ node—such as CPU load, energy consumption, or bandwidth usage—and $n$ is the total number of nodes. The index produces a value between 0 and 1, where 1 indicates perfect no bottlenecks, and values closer to 0 indicate increasing bottlenecks. Several fairness index variants based on Jain’s approach have been proposed; \textit{Donta et al.} \cite{8882255} offers an alternate formulation.

\subsubsection{Measure Observability}  by making the processes and interactions within the DCC transparent and understandable, explainability empowers real-time monitoring, debugging, and continuous improvement, ensuring that complex, heterogeneous, and autonomous systems remain accountable, reliable, and adaptable. Explainability enhances observability by enabling stakeholders to move beyond basic monitoring and gain meaningful insights into the rationale behind automated system behaviors and resource allocation decisions. This interpretability allows operators to trace the origins of actions and outcomes across cloud, edge, and IoT nodes, facilitating more effective diagnosis of issues and optimization of system performance. The observability score for a DCC is calculated as Eq.~(\ref{eq_Observe}).
\begin{equation}\label{eq_Observe}
\mathcal{E} = \frac{ \sum_{i=1}^{N} \gamma_i \, E_{\text{local}}^i }{ N } \cdot \left( 1 - \frac{ \sum_{i \neq j} |C_{ij} - C_{ji}| }{ \sum_{i,j} C_{ij}} \right)
\end{equation}
where, $N$ is the total number of nodes, $E_{\text{local}}^i$ is the local explainability score of node $i$, and $\gamma_i$ is the weight for node $i$. $C_{ij}$ represents causal influence from node $i$ to node $j$, which can be estimated using techniques such as Granger causality or other causal inference methods\cite{pujol2024causality}. In short, the first term in Eq.~(\ref{eq_Observe}) captures the average explainability across nodes, while the second term penalizes asymmetric or unclear inter-node interactions. 

\subsubsection{Adaptivity Quotient} is a metric designed to quantify how effectively and rapidly a DCC system can sense, respond to, and recover from changes or disruptions across the continuum. In DCC environments, adaptation is not limited to simple resource scaling; it also encompasses dynamic workload migration, service offloading, quality adjustment, and reconfiguration in response to fluctuating workloads, failures, or environmental changes. As recent research introduces adaptive frameworks, the ability to adapt is crucial for ensuring service continuity, meeting application requirements, and optimizing resource use across the continuum. This metric is mathematically defined as shown in Eq.(\ref{eq_AQ}).
\begin{equation}\label{eq_AQ}
\mathcal{Q} = \frac{1}{K} \sum_{k=1}^{K} \left( \frac{P^{\text{post}}_k}{P^{\text{base}}_k} \cdot \frac{1}{T^{\text{adapt}}_k} \right)
\end{equation}
where, $K$ represents the total number of adaptation events observed in DCC. $P_k^{\text{before}}$ is the  metric (such as throughput or latency) before adaptation event $k$, and $P_k^{\text{after}}$ is the  metric after the system has adapted. $T_k$ is the time taken to complete the adaptation for event $k$.

\subsubsection{Maintain Equilibrium} 
Scaling up computational resources comes at a \textbf{cost}, leading to higher operational expenses, increased energy demands (projected to consume 7\% of global power by 2030), and greater \textit{environmental impact} \cite{chen2025data}. While investing in infrastructure enhances QoS—ensuring lower latency and better reliability—it also amplifies resource consumption and financial overhead. For instance, increasing replication factors to improve data availability demands additional storage and processing power, while low-latency techniques like indexing and caching require more computational resources. Conversely, reducing computational power or storage capacity affects QoS, creating a self-perpetuating challenge: more resources improve performance but increase costs, while cost-cutting compromises system reliability. These investments raise the question of how effectively increased resources translate into real-world performance gains. Notably, the growth of \textbf{personalized LLMs} may not require such large computing environments but necessitates careful control over infrastructure due to the associated costs, especially for small organizations (e.g., universities or public services). For example, Amdahl’s Law (Eq.~(\ref{eqAmdhals}), where the variable $S$ represents speedup and $F$ represents fraction), which predicted this situation \cite{amdahl2013computer}, highlights the limitation of diminishing returns: performance gains are constrained by the fraction of the workload that remains unoptimized. Therefore, the challenge for large-scale AI infrastructure initiatives lies not just in scaling hardware, but in efficiently using available resources to maximize QoS.
\begin{equation}\label{eqAmdhals}
    S_{overall} = {\left(1-F_{enhanced} + \frac{F_{enhanced}}{S_{enhanced}}\right)^{-1}}
\end{equation}
The interplay between Cost ($C$), Resources ($R$), and QoS ($Q$) creates a complex optimization challenge. Increasing resources enhances QoS but raises costs, while reducing costs can degrade performance. Eq.~(\ref{eq_RCQ}) formalizes this trade-off, aiming QoS while adhering to cost and resource constraints.
\begin{equation}\label{eq_RCQ}
    \max_{R,C} Q(R,C) \text{ subjected to: } C\leq C_{max} \text{ , } R \geq R_{min}
\end{equation}
Current AI-driven infrastructure management lacks adaptive frameworks that can dynamically respond to workload fluctuations while maintaining \textbf{equilibrium}. Existing methods often rely on static provisioning or reactive scaling, leading to inefficiencies. In this context, further research should focus on AI-driven, self-optimizing platform that continuously learns from system fluctuations to achieve near-optimal equilibrium across cost, resources, and QoS, satisfying Eq.(\ref{eq_RCQ}). 

\subsection{Metrics Selection Criteria}
% DCC or IoT performance can be described using the metrics shown in section~\ref{sec3} and subsection~\ref{section41}. However, 
In general, it is not necessary to evaluate all of them in every case. Deciding on appropriate or effective performance metrics is challenging, however, some characteristics can make this task easier. Next, we provide helpful guidelines to decide the set of metrics that best characterize the system. Ideally, the metric should be related to DCC objectives and goals. It is important that the metric be \textit{sensitive} enough to detect changes in the system's performance over time. Even minor behavior changes should be able to be detected by it. In order to be \textit{consistent}, the units of the performance metric and its precise definition must be the same across different systems or configurations of the same system. A performance metric is \textit{repeatable} if it is measured the same way each time an experiment is conducted. An easy-to-understand and \textit{easy-to-use} metric is essential and maintain the results should be clear and concise. All aspects of DCC performance should be captured by the metric. In other words, it needs to take into account all relevant aspects of DCC behavior, such as resource utilization, scalability, reliability, and energy efficiency. Performance metrics should be \textit{independent} of outside influences. When all these requirements are met, a performance metric can be easily determined as to its accuracy and relevance are considered.

\section{Conclusions}\label{sec5}
This paper has explored currently used performance measurements within AI-centric DCC systems across computing, network, and user-centric metrics, highlighting their foundational role. These metrics often fall short in addressing the continuously changing computational demands and application requirements, particularly in the context of GenAI and LLM workloads. To bridge this gap, we propose a set of novel performance metrics—including $CO_2$ emissions, heat dissipation, bottleneck fairness, observability, adaptivity quotient, and equilibrium maintenance—that collectively enable a more holistic and nuanced evaluation of DCC systems. These metrics incorporate critical dimensions such as sustainability, energy efficiency, adaptive responsiveness, and system transparency, which are essential for guiding the optimization of resource use and service quality in DCC. In addition, we outline criteria for selecting appropriate metrics based on relevance, sensitivity, and independence, supporting researchers in aligning measurement practices with operational goals. This paper aims to inspire the adoption of comprehensive, forward-looking performance measurements that will foster the development of efficient, robust, and sustainable AI-driven DCC systems capable of meeting the needs of next-generation applications. However, there remains significant scope for developing additional metrics for future applications; we plan to introduce several other measurements in the near future.

\section*{Acknowledgment}
Many thanks to \textit{Alaa Saleh}, PhD student at University of Oulu, Finland for helping to draw Figures in this paper. %This project is partially supported by the \textit{Svenska Institutet} under the 'SI Baltic Sea Neighbourhood Programme 2024' (Project No. 31005669).
\bibliographystyle{IEEEtran}
\bibliography{ref.bib}
\balance
\end{document}